\documentclass[fleqn,twoside]{article}
\usepackage{espcrc2}

\usepackage{graphicx}
\usepackage[figuresright]{rotating}

\newcommand{\AmS}{{\protect\the\textfont2
  A\kern-.1667em\lower.5ex\hbox{M}\kern-.125emS}}

\title{Structures and electromagnetic properties of the $A$-site ordered/disordered manganites; 
$R$BaMn$_2$O$_6$/$R$$_{0.5}$Ba$_{0.5}$MnO$_3$ ($R$ = Y and rare earth elements)}

\author{T. Nakajima\address{Materials Design and Characterization Laboratory, Institute for Solid State Physics, University of Tokyo, 5-1-5 Kashiwanoha, Kashiwa, Chiba 277-8581, Japan}, Y. Ueda
        }
       
\begin{document}

\begin{abstract}
The discovery of novel structural and physical properties in the $A$-site ordered manganite $R$BaMn$_2$O$_6$ ($R$ = Y and rare earth elements) has demanded new comprehension about perovskite manganese oxides. In the present study, the $A$-site disordered form $R$$_{0.5}$Ba$_{0.5}$MnO$_3$ was successfully synthesized and compared with $R$BaMn$_2$O$_6$ in the structures and electromagnetic properties. $R$$_{0.5}$Ba$_{0.5}$MnO$_3$ has a primitive cubic perovskite cell. Unexpectedly the magnetic glassy states are dominant as the ground state for $R$$_{0.5}$Ba$_{0.5}$MnO$_3$. A peculiar behavior, the steplike ultrasharp magnetization and resistivity change, has been observed in Pr$_{0.5}$Ba$_{0.5}$MnO$_3$.
\vspace{1pc}
\end{abstract}

\maketitle

\section{Introduction}
\begin{figure}[ht]
\begin{center}
\includegraphics[scale=0.8]{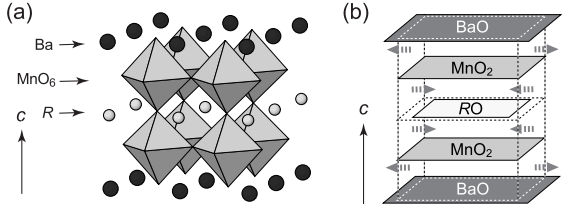}
\end{center}
\caption{(a) Crystal structure and (b) structural concept of the $A$-site ordered manganite $R$BaMn$_2$O$_6$.}
\label{1}
\end{figure}

\begin{figure}[ht]
\begin{center}
\includegraphics[scale=0.6]{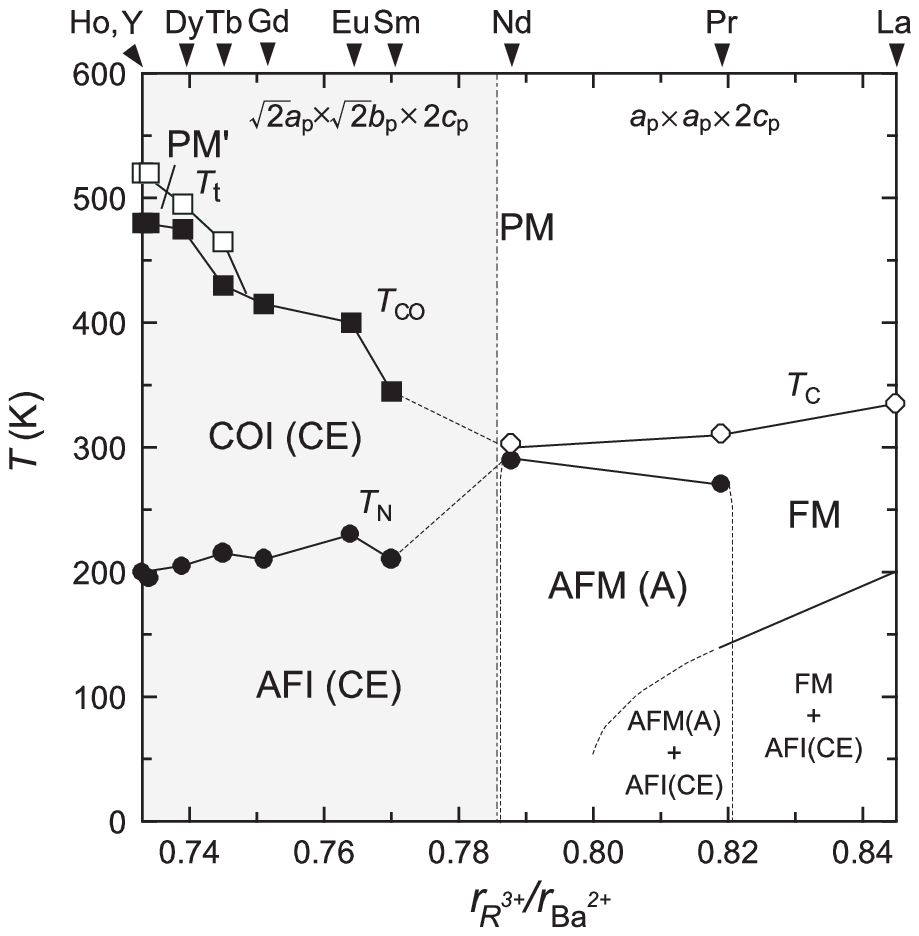}
\end{center}
\caption{Electronic phase diagram of $R$BaMn$_2$O$_6$.}
\label{2}
\end{figure}

The magnetic and electrical properties of perovskite manganites with the general formula ($R^{3 + }_{1 - x}A^{2 + }_{x})$MnO$_{3}$ ($R$ = rare earth elements and $A$ = Sr and Ca) have been extensively investigated for the last decade \cite{1}. Among the interesting features are the so-called colossal magnetoresistance (CMR) and metal-insulator (MI) transition accompanied by charge/orbital order (CO). It is now widely accepted that these enchanting phenomena are caused by the strong correlation/competition of multi-degrees of freedom, that is, spin, charge, orbital and lattice, which could be significantly influenced by the $A$-site randomness. Recently we successfully synthesized the $A$-site ordered manganite $R$BaMn$_2$O$_6$ and reported its structure and electromagnetic properties. As schematically shown in Fig. \ref{1}, the most significant structural feature of $R$BaMn$_2$O$_6$ is that the MnO$_2$ square sublattice is sandwiched by two types of rock-salt layers, $R$O and BaO, with much different sizes and consequently the MnO$_6$ octahedron itself is distorted in a curious manner, in contrast to the rigid MnO$_6$ for the $A$-site disordered manganite ($R_{1 - x}A_{x})$MnO$_{3}$. This means that the structural and physical properties of $R$BaMn$_2$O$_6$ can be no longer explained in terms of the basic structural distortion, the so-called tolerance factor $f$, as in the $A$-site disordered manganites $R$$_{0.5}$Ba$_{0.5}$MnO$_3$. Figure \ref{2} shows the electronic phase diagram of $R$BaMn$_2$O$_6$ expressed as a function of the ratio of ionic radius of the $A$-site cations, $r$$_{R^{3 + }}$/$r$$_{\rm {Ba}^{2 + }}$ [3]. Among possible combinations of $R$/Ba, the mismatch between $R$O and BaO is the smallest in La/Ba and the largest in Y/Ba. When $R^{3 + }$ is smaller than Sm$^{3 + }$ in ion size, the CE-type charge/orbital ordered state (COI(CE)) with a new stacking variation along the $c$-axis is stabilized at the relatively high temperatures ($T_{\rm CO}$) far above 300 K, which would be not only due to the absence of $A$-site randomness but also due to the tilt of MnO$_6$ octahedra as well as heavy distortions of MnO$_6$ octahedron [2-5]. Especially $R$BaMn$_2$O$_6$ ($R$ = Tb, Dy, Ho and Y) shows the structural transition above $T_{\rm CO}$, which is possibly accompanied by the $d_{x^{2} - y^{2}}$ type orbital order [2,3,5]. Therefore the orbital, charge and spin degrees of freedom is separated in these compounds. On the other hand, $R$BaMn$_2$O$_6$ ($R$ = La, Pr and Nd) with relatively larger $R^{3 + }$ has no octahedral tilt and shows a transition from a paramagnetic metal (PM) to a ferromagnetic metal (FM). The ground states for Pr- and Nd-compounds are $A$-type antiferromagnetic metal (AFM(A)). In LaBaMn$_2$O$_6$, the antiferromagnetic CE-type charge and orbital ordered state (AFI(CE)) coexists with FM as the ground state, which suggests that the electronic phase separation is not due to the $A$-site randomness but is intrinsic phenomenon for perovskite manganites. 

Such discovery of novel structural and physical properties in the $A$-site ordered manganite $R$BaMn$_2$O$_6$ has demanded new comprehension about perovskite manganese oxides. However the $A$-site disordered form with the same constituent elements is crucial for such comprehension, especially in order to study the effect of the $A$-site randomness at least qualitatively. We successfully synthesized the $A$-site disordered form $R$$_{0.5}$Ba$_{0.5}$MnO$_3$. In this paper, we report the structure and electromagnetic properties of $R$$_{0.5}$Ba$_{0.5}$MnO$_3$, and compare $R$$_{0.5}$Ba$_{0.5}$MnO$_3$ with both $R$BaMn$_2$O$_6$ and $R$$_{0.5}$$A$$_{0.5}$MnO$_3$.

\section{Experimental}

Powder samples of $R$$_{0.5}$Ba$_{0.5}$MnO$_3$ were prepared by a solid-state reaction of $R$$_2$O$_3$, BaCO$_3$ and MnO$_2$ at 1350\r {}C in 1$\%$ O$_2$/Ar gas, followed by annealing at 900\r {}C in O$_2$ gas for 1 day. Annealing of $R$BaMn$_2$O$_6$ under O$_2$ gas at high temperatures results in insufficient $R$/Ba solid-solution. The obtained products were checked to be of single phase by X-ray diffraction. The degree of $R$/Ba solid-solution at the $A$-site was carefully checked by measuring the intensity of (00$\frac{1}{2}$)$\rm_{p}$ reflection indexed in the primitive cell. We concluded the perfect disorder or perfect solid solution of $R$/Ba only in the case of the absence of (00$\frac{1}{2}$)$\rm_{p}$ reflection.

The crystal structures and cell parameters were determined by X-ray powder diffraction. The magnetic properties were studied using a SQUID magnetometer in a temperature range $T$ = 2$\sim$400 K. The electric resistivity of a sintered pellet was measured for $T$ = 2$\sim$400 K by a conventional four-probe technique.

\section{Results and discussion}

\begin{figure}[ht]
\begin{center}
\includegraphics[scale=0.65]{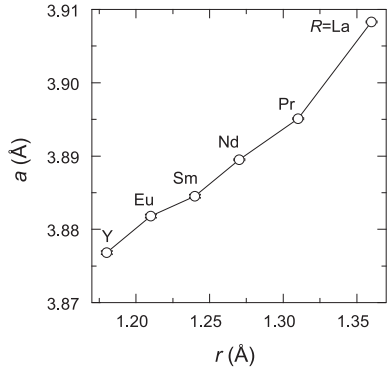}
\end{center}
\caption{Lattice parameters of the $A$-site disordered manganite $R$$_{0.5}$Ba$_{0.5}$MnO$_3$.}
\label{3}
\end{figure}

The X-ray diffraction patterns of all $R$$_{0.5}$Ba$_{0.5}$-MnO$_3$ can be indexed in the primitive cubic perovskite cell. This means no tilt of MnO$_6$ octahedra in contrast to the GdFeO$_3$ type distortion due to the tilt of MnO$_6$ octahedra in $R$$_{0.5}$$A$$_{0.5}$MnO$_3$. In general, the mismatch between the larger MnO$_2$ and the smaller $A$O sublattices is relaxed by tilting MnO$_6$ octahedra, resulting in the lattice distortion from cubic to, mostly, orthorhombic GdFeO$_3$-type structure. The degree of mismatch is described as the tolerance factor $f$ = ($<$$r_{A}$$>$+$r_{\rm O})$/[$\sqrt{2}$($r_{\rm Mn}+r_{\rm O})$], where $<$$r_{A}$$>$, $r_{\rm Mn}$ and $r_{\rm O}$ are (averaged) ionic radii for the respective elements. In $R$$_{0.5}$Ba$_{0.5}$MnO$_3$, the $f$ is in the range from 1.026 (La/Ba) to 0.995 (Y/Ba), which are rather close to $f$ = 1, comparing to the variation 0.955 $<$ $f$ $<$ 1 in $R$$_{0.5}$$A$$_{0.5}$MnO$_3$. The simple cubic structures for $R$$_{0.5}$Ba$_{0.5}$MnO$_3$ can be understood to be partly due to the $f$ values close to 1. The lattice parameters of $R$$_{0.5}$Ba$_{0.5}$MnO$_3$ at room temperature are shown in Fig. \ref{3}. The lattice parameter decreases with decreasing ionic radius of $R$$^{3+}$.

\begin{figure}[ht]
\begin{center}
\includegraphics[scale=0.75]{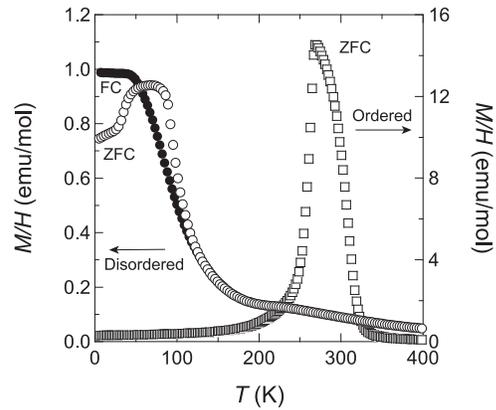}
\end{center}
\caption{Temperature dependence of magnetic susceptibility for the $A$-site ordered/disordered PrBaMn$_2$O$_6$/Pr$_{0.5}$Ba$_{0.5}$MnO$_3$.}
\label{4}
\end{figure}

\begin{figure}[ht]
\begin{center}
\includegraphics[scale=0.75]{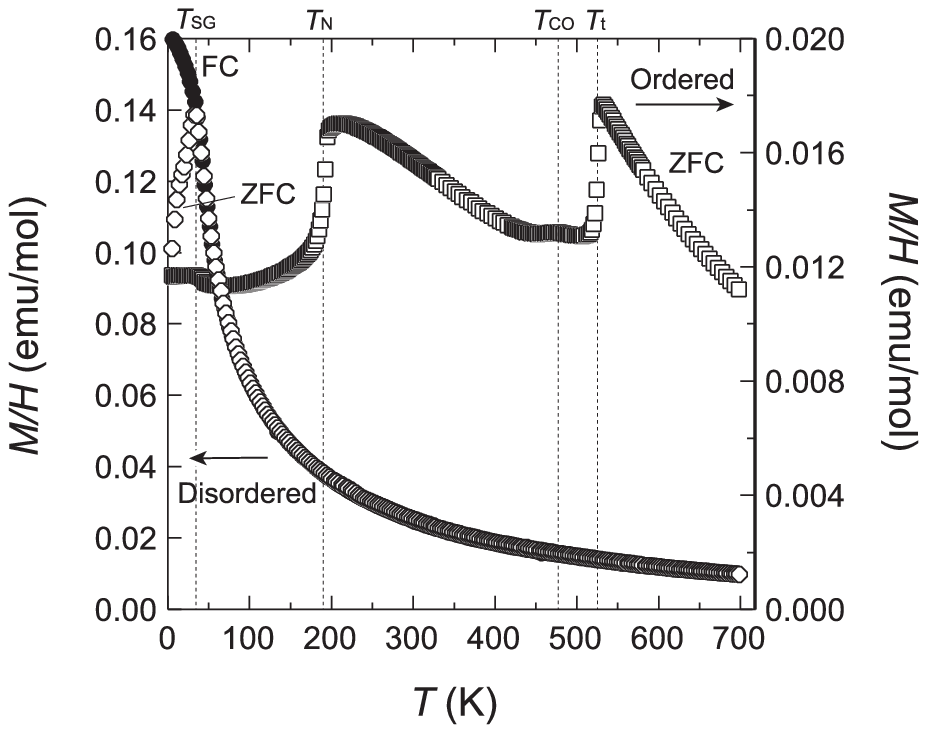}
\end{center}
\caption{Temperature dependence of magnetic susceptibility for the $A$-site ordered/disordered YBaMn$_2$O$_6$/Y$_{0.5}$Ba$_{0.5}$MnO$_3$.}
\label{5}
\end{figure}

\begin{figure}[ht]
\begin{center}
\includegraphics[scale=0.7]{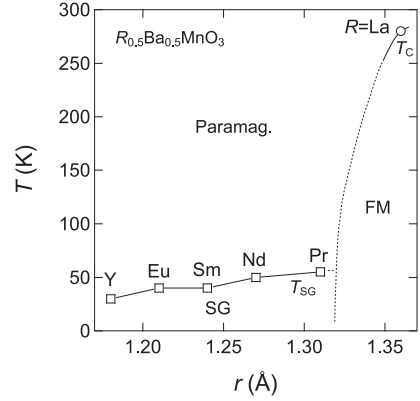}
\end{center}
\caption{Magnetic phase diagram of $R$$_{0.5}$Ba$_{0.5}$MnO$_3$.}
\label{6}
\end{figure}

From the cubic structures of $R$$_{0.5}$Ba$_{0.5}$MnO$_3$, one may expect FM generated by double exchange interaction as the stable electronic state. The ground state of La$_{0.5}$Ba$_{0.5}$MnO$_3$ is actually a pure FM and the TC decreases by 50 K compared with $T_{\rm C}$ = 330 K in LaBaMn$_2$O$_6$, agreeing with the previous report [1]. On the other hand, Pr$_{0.5}$Ba$_{0.5}$MnO$_3$ and Nd$_{0.5}$Ba$_{0.5}$MnO$_3$ have magnetic transitions at $T_{\rm m}$ = 160 K and 120 K, respectively, and show glassy behaviors evidenced by significant differences of magnetic susceptibility ($M/H$)-temperature ($T$) curves on zero-field cooled (ZFC) and field cooled (FC) processes, as shown in Fig. \ref{4}. More typical spin-glass behaviors have been observed in $R$$_{0.5}$Ba$_{0.5}$MnO$_3$ with Sm$^{3+}$ and smaller $R$$^{3+}$s. The typical example of Y$_{0.5}$Ba$_{0.5}$MnO$_3$ is shown in Fig. \ref{5} together with $M/H$ for YBaMn$_2$O$_6$. YBaMn$_2$O$_6$ shows three successive transitions; structural transition at $T_{\rm t}$, CO transition at $T_{\rm CO}$ and antiferromagnetic transition at $T_{\rm N}$. The magnetic interaction is ferromagnetic above $T_{\rm t}$, while it is antiferromagnetic below $T_{\rm t}$. In Y$_{0.5}$Ba$_{0.5}$MnO$_3$, on the other hand, there is no evidence for any transition or no trace of the transitions observed in YBaMn$_2$O$_6$, except for the spin-glass transition at 30 K. The results are shown in Fig. \ref{6} as the phase diagram. Comparing with the phase diagrams of $R$$_{0.5}$Ba$_{0.5}$MnO$_3$ and $R$BaMn$_2$O$_6$, the electronic states characteristic of perovskite manganites such as AFM(A) and COI(CE) are absent in $R$$_{0.5}$Ba$_{0.5}$MnO$_3$. In stead of these states, magnetic glassy states govern the electronic state of $R$$_{0.5}$Ba$_{0.5}$MnO$_3$.

The magnetic glassy state could be due to a disorder effect that hinders the long-range magnetic ordering and could occur as a result of the competition between randomly distributed ferromagnetic and antiferromagnetic interactions. Here it should be emphasized that the glassy states has never observed in ordinary $R$$_{0.5}$$A$$_{0.5}$MnO$_3$. Since the ionic radius of Ba$^{2+}$ is much larger than that of Sr$^{2+}$ (= 1.44 \AA ) and of course $R$$^{3+}$ ($\leq $ 1.36 \AA  ) [9], $R$$_{0.5}$Ba$_{0.5}$MnO$_3$ may include the spatial heterogeneity in nanometer size, which leads to the magnetic nonhomogeneous state. Only in the case of the largest La$^{3+}$ among $R$$^{3+}$s, a homogeneous solid-solution with Ba$^{2+}$ is formed in the $A$-site of La$_{0.5}$Ba$_{0.5}$MnO$_3$ as in the case of $R$$_{0.5}$$A$$_{0.5}$MnO$_3$, and the long range magnetic ordering, FM, is realized. Here, it should be noticed that the magnetic state of $R$$_{0.5}$Ba$_{0.5}$MnO$_3$ was significantly affected by the degree of the $A$-site disorder. The $R$$_{0.5}$Ba$_{0.5}$MnO$_3$ ($R$ = La, Pr and Nd) prepared from annealing $R$BaMn$_2$O$_6$ has insufficient disorder of the $A$-site and shows the FM transition with monotonous decrease of $T_{\rm C}$ as La $>$ Pr $>$ Nd [13].

\begin{figure}[ht]
\begin{center}
\includegraphics[scale=0.8]{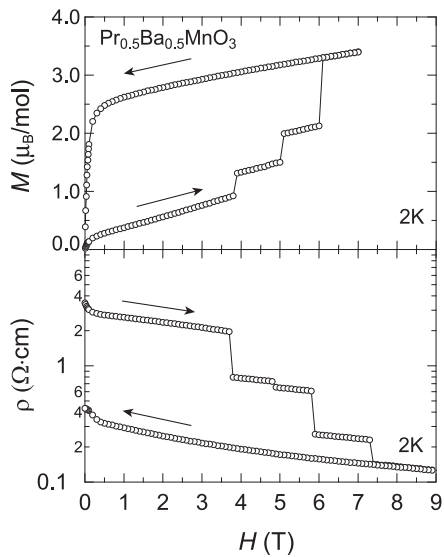}
\end{center}
\caption{Magnetic field dependences of magnetization and resistivity at 2K for  Pr$_{0.5}$Ba$_{0.5}$MnO$_3$.}
\label{7}
\end{figure}

Very interestingly, a peculiar behavior has been observed in Pr$_{0.5}$Ba$_{0.5}$MnO$_3$ at 2 K, as shown in Fig. 7. The resistivity decreases stepwise as the magnetic field increases, while the magnetization increases stepwise with the close relation to the resistivity behaviors. These behaviors are not reversible in the magnetic field. The stepwise behaviors in the magnetization and resistivity were observed up to 4.9 K but they vanished dramatically at 5.0 K. Similar behaviors were previously reported in Pr$_{0.5}$Ca$_{0.5}$MnO$_3$ doped with a few percent of other cations such as Sc, Ga or Co on the Mn site and were explained by an impurity induced-disorder, with the coexistence of several short-range AFI(CE) phases and small FM regions [14,15]. Our system has neither FM-to- AFI(CE) transition nor dopant in contrast to Pr$_{0.5}$Ca$_{0.5}$MnO$_3$. This is the first observation of ultrasharp magnetization and resistivity change in the nondoped system. A model based on ordinal two-phase mixture cannot explain the behavior. For instance, the AFI(CE) phase in the coexistence with the FM phase is continuously converted to FM phase as observed in LaBaMn$_2$O$_6$. We have no explanation for such ultrasharp magnetization and resistivity change at present. However we would like to emphasize a close relation between the observed behavior and the special heterogeneity in nanometer size. Detailed study is now in progress.

To summarize, we have investigated the structures and electromagnetic properties of the $A$-site ordered/disordered manganite $R$BaMn$_2$O$_6$/$R$$_{0.5}$-Ba$_{0.5}$MnO$_3$. The disordered form $R$$_{0.5}$Ba$_{0.5}$MnO$_3$ has primitive cubic perovskite cell with no tilt of MnO$_6$ octahedra. Unexpectedly the magnetic glassy states are dominant as the ground state for $R$$_{0.5}$Ba$_{0.5}$MnO$_3$. Since the ionic radius of Ba$^{2+}$ is much larger than that of Sr$^{2+}$ (= 1.44 \AA ) [9], $R$$_{0.5}$Ba$_{0.5}$MnO$_3$ may include the special heterogeneity in nanometer size, which leads to the magnetic nonhomogeneous state. The magnetic glassy state could be due to a disorder effect that hinders the long-range magnetic ordering and could occur as a result of the competition between randomly distributed ferromagnetic and antiferromagnetic interactions. As the remarkable phenomena of $R$$_{0.5}$Ba$_{0.5}$MnO$_3$, the steplike ultrasharp magnetization and resistivity changes have been observed in Pr$_{0.5}$Ba$_{0.5}$MnO$_3$.
The authors thank H. Kageyama, T. Yamauchi, M. Isobe, Y. Matsushita, Z. Hiroi, H. Fukuyama and K. Ueda for valuable discussion. This work is partly supported by Grants-in-Aid for Scientific Research (No. 407 and No. 758) and for Creative Scientific Research (No. 13NP0201) from the Ministry of Education, Culture, Sports, Science, and Technology.




\begin{thebibliography}{00}


\bibitem{1} F. Millange, V. Caignaert, B. Domeng\`{e}s and B. Raveau, Chem. 
Mater. {\bf 10} (1998) 1974.

\bibitem{2} T. Nakajima, H. Kageyama and Y. Ueda, J. Phys. Chem. Solids {\bf 63} (2002) 913.

\bibitem{3} T. Nakajima, H. Kageyama, H. Yoshizawa and Y. Ueda, J. Phys. Soc. Jpn. {\bf 71} (2002) 2843.

\bibitem{4} H. Kageyama, T. Nakajima, M. Ichihara, Y. Ueda, H. Yoshizawa, and K. Ohoyama, J. Phys. Soc. Jpn. {\bf 72} (2003) 241.

\bibitem{5} T. Nakajima, H. Kageyama, K. Ohoyama, H. Yoshizawa, and Y. Ueda, submitted to J. Solid State Chem.

\bibitem{6} S. V. Trukhanov, I. O. Troyanchuk, M. Hervieu, H. Szymczak and K. Barner, Phys. Rev. B {\bf 66} (2002) 184424.

\bibitem{7} M. Uchida, D. Akahoshi, R. Kumai, Y. Tomioka, T. Arima, Y. Tokura, and Y. Matsui, J. Phys. Soc. Jpn. {\bf 71} (2002) 2605.

\bibitem{8} T. Arima, D. Akahoshi, K. Oikawa, T. Kamiyama, M. Uchida, Y. Matsui, and Y. Tokura, Phys. Rev. B {\bf 66} (2002) 140408.

\bibitem{9} R. D. Shannon, Acta Crystallogr. A{\bf 32} (1976) 751.

\bibitem{10} F. Izumi and T. Ikeda, Mater. Sci. Forum {\bf 321-324} (2000) 198.

\bibitem{11} Y. Tomioka, A. Asamitsu, H. Kuwahara, Y. Moritomo and Y. Tokura, Phys. Rev. B {\bf 53} (1996) 1689.

\bibitem{12} See for reviews: C. N. R. Rao and B. Raveau, \textit{Colossal Magnetoresistance, Charge Ordering and Related Properties of Manganese Oxides}; World Scientific, Singapore (1998).

\bibitem{13} T. Nakajima $et$ $al$., in preparation

\bibitem{14} S. H\`{e}bert, V. Hardy, A. Maignan, R. Mahendiran, M. Hervieu, C. Martin and B. Raveau, J. Solid State Chemistry {\bf 165} (2002) 6.

\bibitem{15} R. Mahendiran, A. Maignan, S. H\`{e}bert, C. Martin, M. Hervieu, B. Raveau, J. F. Mitchell and P. Schiffer, Phys. Rev. Lett., {\bf 89} (2002) 286602.
 
\end{thebibliography}
\end{document}